\documentclass[twocolumn]{aastex63}

\usepackage{amsmath,amssymb,lineno}
\usepackage{tikz-feynman}
\usepackage{newtxmath}

\submitjournal{ApJ}

\shorttitle{}
\shortauthors{}

\newcommand{\dif}{\mathrm{d}}

\begin{document}

\title{Transport of charged particles propagating in turbulent magnetic fields as a red-noise process}
\email{deligny@ijclab.in2p3.fr}

\author[0000-0001-6863-6572]{Olivier Deligny}
\affiliation{Laboratoire de Physique des 2 Infinis Ir\`ene Joliot-Curie\\
CNRS/IN2P3, Universit\'{e} Paris-Saclay, Orsay, France}

\begin{abstract}
The transport of charged particles in various astrophysical environments permeated by magnetic fields is described in terms of a diffusion process, which relies on diffusion-tensor parameters generally inferred from Monte-Carlo simulations. In this paper, a theoretical derivation of the diffusion coefficient in the case of a purely turbulent magnetic field is presented. The approach is based on a red-noise approximation to model the 2-pt correlation function of the magnetic field experienced by the particles between two successive times. This approach is shown to describe the regime in which the Larmor radius of the particles is in resonance with the wavelength power spectrum of the turbulence (gyro-resonant regime), extending hence previous results applying to the high-rigidity regime in which the Larmor radius is greater than the larger wavelength of the turbulence. The results are shown to be consistent with those obtained with a Monte-Carlo generator. Although not considered in this study, the presence of a mean field on top of the turbulence is discussed.   
\end{abstract}



\section{Introduction} 
\label{sec:intro}

Many astrophysical environments such as jets, galaxies, clusters of galaxies, and interplanetary, interstellar or intergalactic space are considered as collisionless turbulent plasmas for the propagation and acceleration of high-energy charged particles (cosmic rays), the confinement and transport of which are governed by their scattering off magnetic turbulence. This is because the fluctuating magnetic field permeating these environments acts as an effective source of collisions in the transport equation of the velocity distribution of the particles~\citep{1972ApJ...172..319J}. Once approximated as a relaxation process, the effective collision term tends to bring the average velocity distribution to its isotropic mean~\citep{PhysRev.94.511}. It is then well established that the flux of particles can be related to their gradient of density by means of a diffusion tensor $D_{ij}$, which can be expressed in terms of the magnetic field unit vector $\mathbf{b}$, the diffusion coefficients parallel and perpendicular to the mean field $D_\parallel$ and $D_\perp$, and the anti-symmetric diffusion coefficient $D_A$ describing the particle drifts as~\citep{1990ApJ...361..162J}
\begin{equation}
\label{eqn:Dij}
    D_{ij}=D_\perp\delta_{ij}+\left(D_\parallel-D_\perp\right)b_ib_j+D_A\epsilon_{ijk}b_k.
\end{equation}

As long as the fluctuating field $\mathbf{\delta B}$ is subdominant with respect to the regular field $\mathbf{B}$, the level of turbulence defined as $\eta=\delta B^2/(\delta B^2+B^2)$ is low and the diffusion coefficients can be determined using a quasi-linear theory approach~\citep{1966ApJ...146..480J,1973ApJ...183.1029J}. However, there are many situations of interest for which the turbulence level is found to be of the order of $0.5$ (turbulent field of the order of the regular field) or even close to $1$ (pure turbulent field). In such turbulence level regimes, many estimates of the diffusion coefficients have been made from numerical simulations exploring wide ranges of particle rigidities~\citep[e.g.][]{1999ApJ...520..204G,PhysRevD.65.023002,Candia:2004yz,2010ApJ...711..997H,Fatuzzo:2014hua,Snodin:2015fza,Reichherzer:2019dmb,Reichherzer:2021yyd}. 

In the high rigidity regime, which is relevant in situations where the Larmor radius of the particles exceeds the coherence length of the turbulence, a theoretical derivation of the coefficients, in agreement with the numerical results, has been formulated in~\cite{Plotnikov:2011me}. It is based on the exact estimation of the average velocity of the particles propagating in the turbulence as a function of time, which is expressed as a Dyson series. In this regime, the values of the magnetic field experienced by the particles decorrelate on time scales much smaller than that of the scattering. This allows the use of a white-noise model for the 2-pt function of the magnetic field experienced between two successive times. Under these conditions, the re-summation of the Dyson series gives rise to an exponential decay of the average velocity characteristic of a Markovian process.

The aim of this study is to extend the results presented in~\cite{Plotnikov:2011me} to a range of  rigidities gyro-resonant with the power spectrum of turbulence. Although the methods employed can be applied to any type of turbulence, this paper is limited, without loss of generalities, to the study of an isotropic 3D turbulence without helicity. The general properties of such a 3D turbulence are reminded in Section~\ref{sec:transport}, as well as the numerical and formal methods based on a Dyson series to determine the time evolution of the mean particle velocities and the 2-pt velocity functions. The red-noise assumption to model the 2-pt function of the magnetic field experienced between two successive times, which allows the introduction of a short time memory, is presented in Section~\ref{sec:acmodel}. It is then shown that this assumption, which makes possible a partial re-summation of the Dyson series, allows the numerical results to be reproduced at the cost, just as in the white-noise case, of a single time parameter related to the turbulence correlation time. From these methods, the diffusion coefficients, which reduced to a single one in the $\eta=1$ studied case, are reproduced in both the gyro-resonant and high-energy regimes. The presence of a mean field on top of a turbulence is discussed in Section~\ref{sec:meanfield}, where some theoretical hints are given. The results are finally discussed in Section~\ref{sec:discussion}.

\section{Transport of charged particles in isotropic magnetic turbulence}
\label{sec:transport}

\subsection{Diffusion of charged particles in magnetic turbulence}
\label{subsec:turbulence}

For a random motion, the spatial diffusion tensor is known to be related under very broad conditions to the velocity correlation function, $\langle v_{0i}v_{j}(t)\rangle$, through a time integration~\citep{Kubo:1957mj}, 
\begin{equation}
\label{eqn:Dij-Kubo}
    D_{ij}(t)=\int_0^{t}\dif t'~\langle v_{0i}v_{j}(t')\rangle,
\end{equation}
in the limit that $t\rightarrow\infty$. Here, $v_{0i}\equiv v_i(t=0)$ and $\langle\cdot\rangle$ stands for the average quantities, taken over several space and time correlation scales of the turbulent field. Throughout the paper, since cosmic rays are high-energy relativistic particles, the norm of the velocity is identified to $c$ for convenience. The fluctuations are considered as ergodic, in the sense that averaging over an ensemble of systems would lead to the same average quantities as through the operation $\langle\cdot\rangle$. The aim of the study is hence to determine a semi-analytical expression for the velocity correlation function. The fluctuating magnetic field, denoted as $\delta\mathbf{B}(\mathbf{x})$ in the spatial space and $\delta\mathbf{B}(\mathbf{k})$ in the reciprocal Fourier one, is characterized in a standard way as a Gaussian random field with zero mean and root mean square value $\delta B^2$ modeling a 3D homogeneous and isotropic turbulence without helicity. For an homogeneous turbulence, the 2-pt correlation function between two components of $\delta\mathbf{B}(\mathbf{x})$ is invariant under spatial translations. In the Fourier space, this translates into 
\begin{equation}
\label{eqn:dBdB}
\langle\delta B_i(\mathbf{k})\delta B^\star_j(\mathbf{k'})\rangle=P_{ij}(\mathbf{k})\delta(\mathbf{k}-\mathbf{k'}),
\end{equation}
which states that two Fourier components of the field are uncorrelated at different wave-number vectors. The $P_{ij}$ quantity is the spectral tensor defined as the Fourier transform of the 2-pt correlation function. 
To guarantee the solenoidal nature of the field, the spectral tensor must satisfy $k_iP_{ij}=k_jP_{ij}=0$, a condition which, combined with the 3D isotropic character of the turbulence imposing rotational invariance on the 2-pt correlation function and with the invariance by symmetry with respect to a plane (no helicity of the field), implies the form $P_{ij}(\mathbf{k})=\beta(k)(\delta_{ij}-k_ik_j/k^2)$~\citep{Batchelor70}, with $\beta(k)$ any function at this stage. On inserting this expression into the kinetic energy spectrum of the turbulence defined as
\begin{equation}
\label{eqn:Ek}
\mathcal{E}(k)=\frac{1}{2}\int_{\Sigma_k}\dif\Sigma~P_{ii}(\mathbf{k}),
\end{equation}
with $\Sigma_k$ standing for the sphere of radius $k$ in the Fourier space, one is left with the relationship $\mathcal{E}(k)=4\pi k^2\beta(k)$ so that the spectral tensor can finally be expressed as a function of a directly interpreteable quantity:
\begin{equation}
\label{eqn:Pij}
P_{ij}(\mathbf{k})=\frac{\mathcal{E}(k)}{4\pi k^2}\left(\delta_{ij}-\frac{k_ik_j}{k^2}\right).
\end{equation}
The size of the largest ``eddies'', $L_{\mathrm{max}}$, is given by the distance over which the correlation function is non-zero. This translates into a minimum wave number $k_\mathrm{min}=2\pi/L_{\mathrm{max}}$ beyond which, for the Kolmogorov turbulence adopted here, the spectrum function follows a power law, $\mathcal{E}(k)=\mathcal{E}_0 k^{-5/3}$, up to a maximum wave number $k_{\mathrm{max}}=2\pi/L_{\mathrm{min}}$, where $L_{\mathrm{min}}$ corresponds to the scale at which the dissipation rate of the turbulence overcomes
the energy cascade rate. The normalisation $\mathcal{E}_0$ is such that $\langle\left|\delta\mathbf{B}(\mathbf{x})\right|^2\rangle=\delta B^2$:
\begin{equation}
\label{eqn:E0}
\mathcal{E}_0=\frac{(2\pi)^{2/3}\delta B^2}{3\left(L_{\mathrm{max}}^{2/3}-L_{\mathrm{min}}^{2/3}\right)}.
\end{equation}
There are several ways to characterize the distance $L_\mathrm{c}$ over which the correlation function of the turbulence is non-zero in the real space. We follow here that of~\cite{Plotnikov:2011me}, 
\begin{equation}
    \label{eqn:Lc}
    L_\mathrm{c}=\int_0^\infty\dif r~\frac{\langle\delta B_i(\mathbf{x})\delta B_i(\mathbf{x+r})\rangle}{\delta B^2},
\end{equation}
which leads, for the Kolmogorov turbulence, to $L_\mathrm{c}=(L_{\mathrm{max}}^{5/3}-L_{\mathrm{min}}^{5/3})/[10(L_{\mathrm{max}}^{2/3}-L_{\mathrm{min}}^{2/3})]$. This is a quantity of interest to estimate in Section~\ref{subsec:dbdb}, the correlation time that provides the duration beyond which a particle experiences a field value decorrelated from the initial one.

\subsection{Monte-Carlo approach}
\label{subsec:mc}

To serve as a reference for testing the model below, 
a Monte-Carlo estimation of the velocity correlation function is used. The strategy of this Monte-Carlo experiment is similar to that widely used in the literature. A large number of particle trajectories in given turbulent magnetic field configurations is simulated by solving numerically the Lorentz-Newton equation of motion that preserves the energy (and hence the Lorentz factor) of the particles:
\begin{equation}
    \label{eqn:LorentzNewton}
    \dot{v}_i(t)=\delta\Omega ~\epsilon_{ijk}v_j(t)\delta b_k(t).
\end{equation}
Here, $\delta\Omega=c^2Z|e|\delta B/E$ is the gyrofrequency with $Z|e|$ the electric charge and $E$ the energy of the particle, and $\delta b_k(t)\equiv\delta b_k(\mathbf{x}(t))$ is the $k$-th component of the magnetic field, expressed in units of $\delta B$, at the spatial coordinate $\mathbf{x}(t)$ of the particle at time $t$. The numerical integration of equation~\ref{eqn:LorentzNewton} is performed using the standard Runge-Kutta integrator. To approximate numerically the isotropic and spatially homogeneous turbulent field, an algorithm similar to that in~\cite{Batchelor70,1999ApJ...520..204G} is used. 
The recipe consists in summing a large number $N_m$ of plane waves ($N_m=250$ in this study) with corresponding wave vector $\mathbf{k}_n$, the direction, phase $\phi_n$ and polarisation of which are chosen randomly:
\begin{equation}
\label{eqn:dBMC}
\delta\mathbf{B}(\mathbf{x})=\sum_{n=1}^{N_m} \sum_{\alpha=1}^{2} \mathcal{E}_n(k_n)~\mathbf{\hat{\xi}}_n^\alpha~\cos{(\mathbf{k}_n\cdot\mathbf{x}+\phi_n^\alpha)}.
\end{equation}
To ensure the condition $\nabla\cdot\delta \mathbf{B}=0$, the two orthogonal polarisation vectors $\mathbf{\hat{\xi}}_n^\alpha$ are oriented in the plane perpendicular to the directions of the wave vectors. The wave number distribution is built from a constant logarithmic spacing between $k_{\mathrm{min}}$ and $k_{\mathrm{max}}$. The wave amplitudes satisfy $\mathcal{E}_n^2(k_n)=\mathcal{E}_0\delta B^2k_n^{-5/3}(k_n-k_{n-1})$, where $\mathcal{E}_0$ is a normalisation factor such that $\sum_n \mathcal{E}_n^2(k_n)=\delta B^2$. The dynamic range of the turbulence explored here is $L_\mathrm{max}/L_\mathrm{min}=100$.

\subsection{Formal approach: Dyson series}
\label{subsec:formal}

Due to the stochastic nature of the magnetic field, the velocity of the particles is a stochastic variable as well, the probability density function of which can be sampled by means of the Monte-Carlo generator described in Section~\ref{subsec:mc}. Formally, the moments of this underlying distribution can be obtained by expressing the solution of equation~\ref{eqn:LorentzNewton} as a Dyson series. The first moment reads as
\begin{eqnarray}
    \label{eqn:dyson}
    \langle v_{i_0}(t)\rangle&=&v_{0i_0}+\sum_{n=1}^\infty \delta\Omega^n ~ \epsilon_{i_0i_1j_1}\epsilon_{i_1i_2j_2}\dots\epsilon_{i_{n-1}i_nj_n} v_{0i_n}\nonumber \\ 
    &\times& \int_0^t \hspace{-0.25cm}\dif t_1\int_0^{t_1}\hspace{-0.35cm}\dif t_2\dots\int_0^{t_{n-1}}\hspace{-0.60cm}\dif t_n \langle\delta b_{j_1}(t_1)\dots\delta b_{j_n}(t_n)\rangle,
\end{eqnarray}
which requires evaluating the expectation value in the integrand of the right-hand side. In the Gaussian approximation, the Wick theorem allows for expressing this expectation value in terms of all possible permutations of products of contractions of pairs of $\langle \delta b_{i_1}(t_{j_1})\delta b_{i_2}(t_{j_2})\rangle$. Using the Ansatz 
\begin{equation}
\label{eqn:dbdb}
\langle \delta b_{i_1}(t_{j_1})\delta b_{i_2}(t_{j_2})\rangle=\frac{\delta_{i_1i_2}}{3}\varphi(t_{j_1}-t_{j_2}),
\end{equation}
and making use of the summation properties over one or two indexes of the Levi-Civita symbol, the first moment of the velocities can be expressed as $\langle v_i(t)\rangle=u(t)v_{0i}$, where the ``propagator'' $u(t)$ reads as
\begin{eqnarray}
    \label{eqn:udyson}
    u(t)&=&1+\sum_{n=1}^\infty  \left(\frac{-2~\delta\Omega^2}{3}\right)^{n} \nonumber \\
    &\times& \int_0^t\hspace{-0.2cm}\dif t_1\int_0^{t_1}\hspace{-0.3cm}\dif t_2\dots\int_0^{t_{2n-1}}\hspace{-0.5cm}\dif t_{2n} \sum_{\{i<j\}}^{}\prod_{} \varphi(t_i-t_j).
\end{eqnarray}
Here, the notation $\sum_{\{i<j\}}^{}\prod_{} \varphi(t_i-t_j)$ stands for the $(2n-1)!!$ permutations of products of contractions of pairs.

In the following, it will be useful to represent the various terms of the expansion of $u(t)$ in the form of diagrams. The function $u(t)$ is considered as a propagator denoted by a double line, while a single line stands for the corresponding ``free propagator'', $u^{(0)}(t)=1$, which can be inserted in between the contraction of a pair, $\langle b_{i_1}(t_{j_1})\delta b_{i_2}(t_{j_2})\rangle = \langle \delta b_{i_1}(t_{j_1})u^{(0)}(t)\delta b_{i_2}(t_{j_2})\rangle$, to build an ``interaction''. A curved dotted line connecting two ``vertices'' then stands for a time-ordered integration over an average product of two stochastic fields:
\begin{equation}\label{eqn:uqlt}
\begin{tikzpicture}
  \begin{feynman}
    \vertex (a1);
    \vertex [right=0.3cm of a1] (a2);
    \vertex [right=0.7cm of a2] (a3);
    \vertex [right=0.3cm of a3] (a4){~=~};
    \diagram*{
      (a1) --[plain] (a4),  
      (a2) -- [scalar, out=90, in=90, looseness=2.0,thick] (a3),
    };
  \end{feynman}
\end{tikzpicture}
\left(\frac{-2\delta\Omega^2}{3}\right)\int_0^t\dif t_1\int_0^{t_1}\dif t_2~\varphi(t_1-t_2).
\end{equation}
The curved dotted line can hence be thought as the propagator of a fictitious field. A comprehensive presentation of the diagrammatic rules can be found in, e.g.,~\cite{Bourret1962,1966AnAp...29..645F,1973JMP....14..531T}. In these conditions, the first terms of the expansion of $u(t)$ can be written as
\begin{equation}\label{eqn:udyson-ex}
\begin{tikzpicture}
  \begin{feynman}
    \vertex (a1);
    \vertex [right=1.0cm of a1] (a2){~=~};
    \vertex [right=0.4cm of a2] (a3);
    \vertex [right=1.0cm of a3] (a4){~+~};
    \vertex [right=0.5cm of a4] (a5);
    \vertex [right=0.2cm of a5] (a6);
    \vertex [right=0.7cm of a6] (a7);
    \vertex [right=0.2cm of a7] (a8){~+~};
    \vertex [right=0.5cm of a8] (a9);
    \vertex [right=0.2cm of a9] (a10);
    \vertex [right=0.7cm of a10] (a11);
    \vertex [right=0.3cm of a11] (a12);
    \vertex [right=0.7cm of a12] (a13);
    \vertex [right=0.2cm of a13] (a14){~+~};
    \vertex [below=of a2] (a15);
    \vertex [right=0.5cm of a15] (a16);
    \vertex [right=0.3cm of a16] (a17);
    \vertex [right=0.5cm of a17] (a18);
    \vertex [right=0.5cm of a18] (a19);
    \vertex [right=0.5cm of a19] (a20);
    \vertex [right=0.3cm of a20] (a21){~+~};
    \vertex [right=0.5cm of a21] (a22);
    \vertex [right=0.3cm of a22] (a23);
    \vertex [right=0.5cm of a23] (a24);
    \vertex [right=0.5cm of a24] (a25);
    \vertex [right=0.5cm of a25] (a26);
    \vertex [right=0.3cm of a26] (a27){~+~...};
    \diagram*{
      (a1) --[double,double distance=0.3ex,thick] (a2),
      (a3) --[plain] (a4),
      (a5) --[plain] (a8),  
      (a6) -- [scalar, out=90, in=90, looseness=2.0,thick] (a7),
      (a9) --[plain] (a14),  
      (a10) -- [scalar, out=90, in=90, looseness=2.0,thick] (a11),
      (a12) -- [scalar, out=90, in=90, looseness=2.0,thick] (a13),
      (a16) --[plain] (a21),  
      (a17) -- [scalar, out=90, in=90, looseness=1.5,thick] (a19),
      (a18) -- [scalar, out=90, in=90, looseness=1.5,thick] (a20),
      (a22) --[plain] (a27),  
      (a23) -- [scalar, out=90, in=90, looseness=1.5,thick] (a26),
      (a24) -- [scalar, out=90, in=90, looseness=2.0,thick] (a25),
    };
  \end{feynman}
\end{tikzpicture}
\end{equation}
By introducing a ``mass operator'' that stands for the sum over all the possible connected diagrams and represented as a blob, equation~\ref{eqn:udyson} can be symbolically written as
\begin{equation}\label{eqn:udyson-bis}
\begin{tikzpicture}
  \begin{feynman}
    \vertex (a1);
    \vertex [right=1.2cm of a1] (a2){~=~};
    \vertex [right=0.4cm of a2] (a3);
    \vertex [right=1.2cm of a3] (a4){~+~};
    \vertex [right=0.4cm of a4] (a5);
    \vertex [right=1.2cm of a5] (a6);
    \node[right=0cm of a6,blob] (a7);
    \vertex [right=1.5cm of a7] (a8){.};
    \diagram*{
      (a1) --[double,double distance=0.3ex,thick] (a2),
      (a3) --[plain] (a4),
      (a5) --[plain] (a6),        
      (a7) --[double,double distance=0.3ex,thick]  (a8)
    };
  \end{feynman}
\end{tikzpicture}
\end{equation}

\section{Red-noise approximation}
\label{sec:acmodel}

\subsection{2-pt function of the experienced magnetic field}
\label{subsec:dbdb}

\begin{figure}[ht]
\centering
\includegraphics[width=\columnwidth]{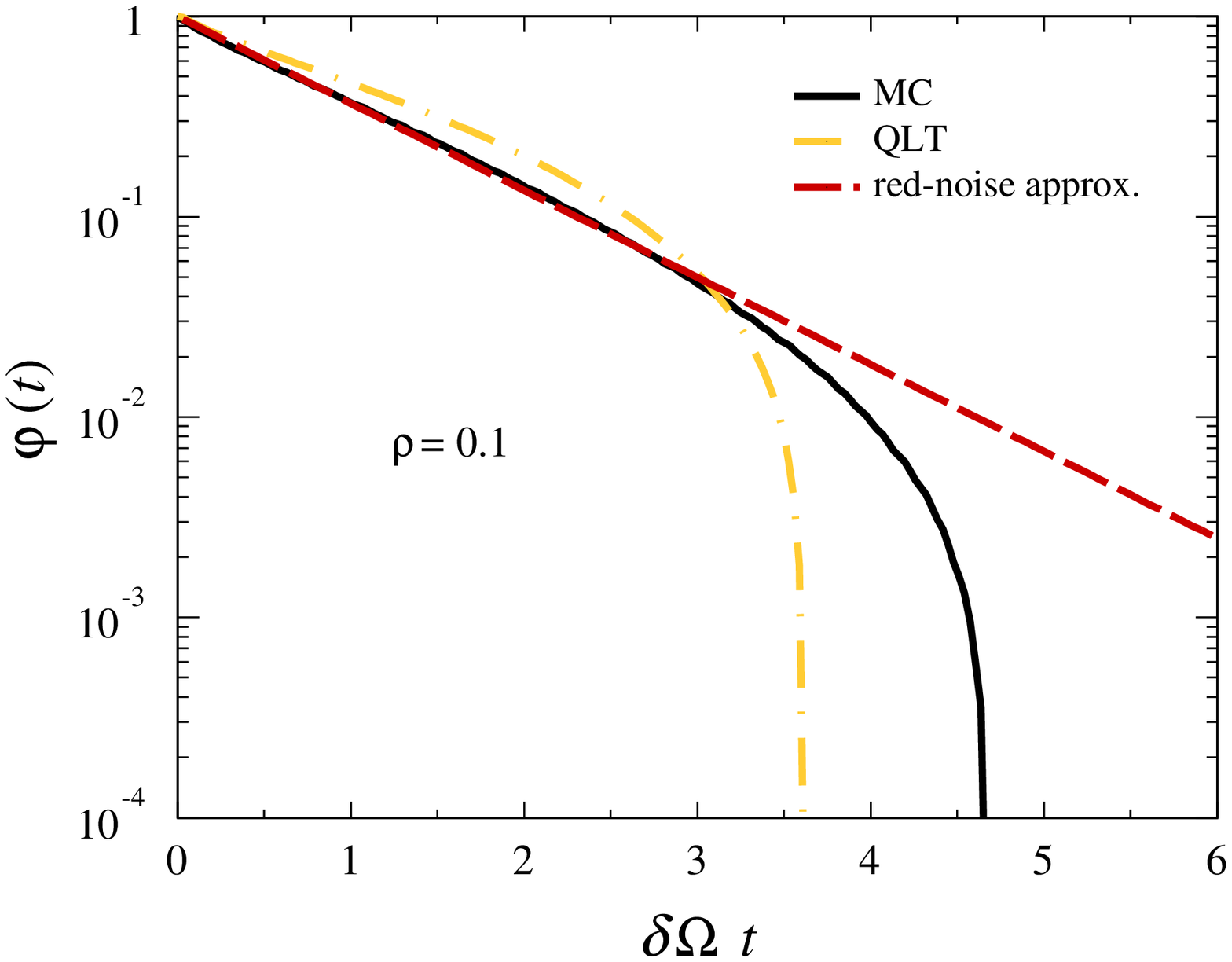}
\includegraphics[width=\columnwidth]{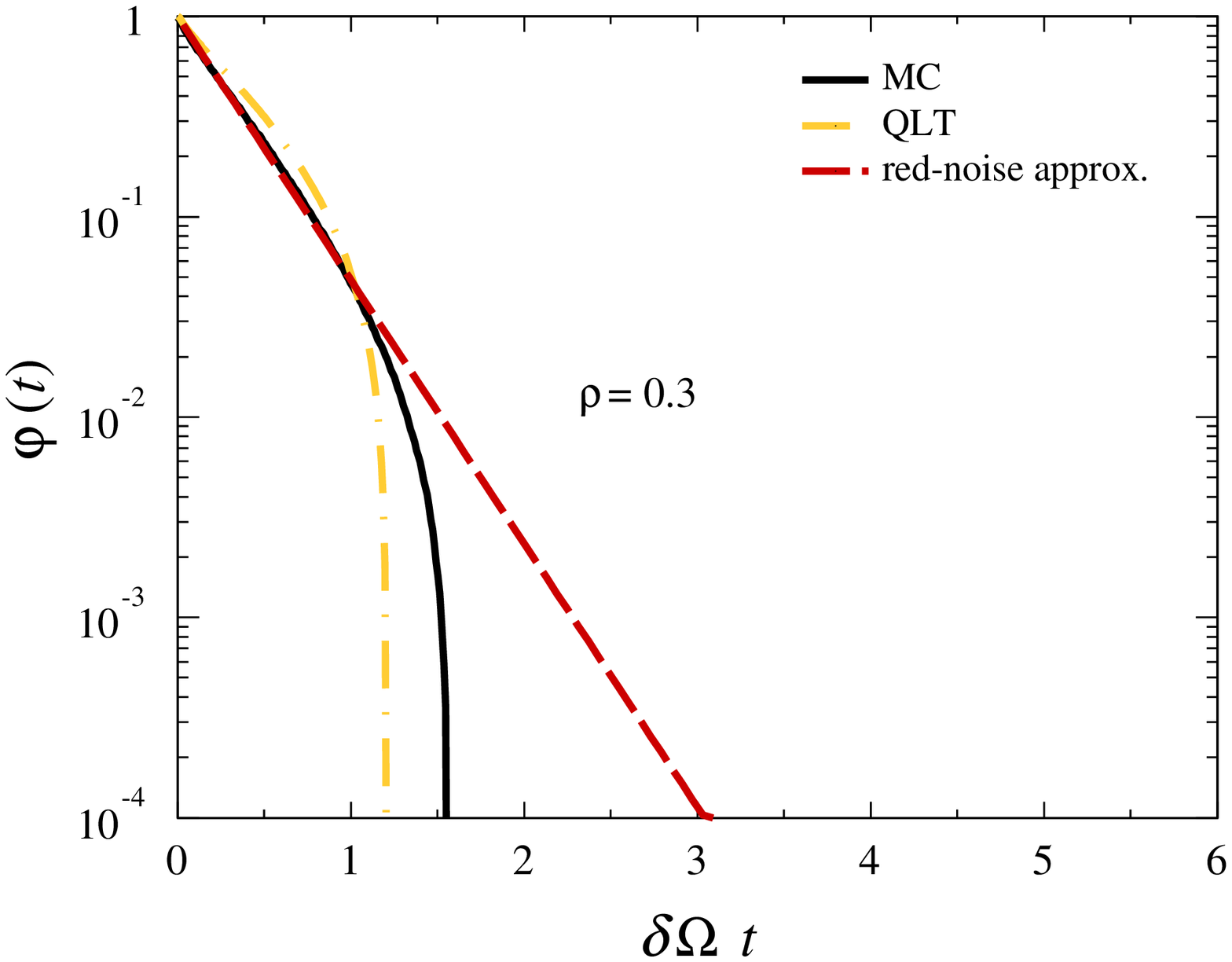}
\caption{Time dependence of the expectation value of $\langle \delta b_i(t)\delta b_i(0)\rangle$ from simulations (MC), quasi-linear theory (QLT) and red-noise approximation. Top: reduced rigidity $\rho=0.1$, bottom: $\rho=0.3$.}
\label{fig:phi}
\end{figure}

A key ingredient to estimate the Dyson series is the Ansatz of the 2-pt correlation function of the magnetic field experienced by the particles between two successive times, which, assuming invariance under time translations, can be explicitly written as
\begin{equation}
    \label{eqn:dbdb_th}
    \langle \delta b_{i}(t)\delta b_{j}(0)\rangle=\iint\dif\mathbf{k}\dif\mathbf{k}'~ \langle \delta b_{i}(\mathbf{k})\delta b_{j}(\mathbf{k}') e^{i\mathbf{k}\cdot\mathbf{x}(t)}\rangle.
\end{equation}
An exact estimation can be made by means of the Monte-Carlo generator. The results obtained in terms of the time-dependent $\varphi(t)$ function are shown in Fig.~\ref{fig:phi} as the black curves for two different reduced rigidities $\rho$, defined as the Larmor radius of the particles expressed in units of $L_\mathrm{max}$, typical of the gyro-resonant regime. For $\rho=0.1$ (top panel), the $\varphi(t)$ function is observed to be non-zero over almost one $2\pi$ period of $\delta\Omega~t$. The non-zero range shrinks while $\rho$ is increasing, as exemplified in the bottom panel with $\rho=0.3$. Eventually, the time scale over which $\varphi(t)$ is non-zero for large $\rho$ values gets so small that a white-noise approximation is accurate~\citep{Plotnikov:2011me}. On the other hand, for rigidities in gyro-resonance with the spectrum of the turbulence, an approximation beyond the white-noise limit is required. 

The simplest approximation beyond the white-noise limit, widely used in the literature, consists in making use first of the independence hypothesis of \cite{Corrsin59} to transform the expectation value in the integrand of the right-hand side of equation~\ref{eqn:dbdb_th} into the product of $\langle \delta b_{i}(\mathbf{k})\delta b_{j}(\mathbf{k}')\rangle$ and $\langle e^{i\mathbf{k}\cdot\mathbf{x}(t)}\rangle$, and second of a quasi-linear theory (QLT) approach to substitute $\mathbf{x}(t)$ for the unperturbed motion $\mathbf{v}_0t$. Averaging over $\mathbf{v}_0$, this yields to the expression
\begin{equation}
    \label{eqn:dbdb_qlt}
    \langle \delta b_{i}(t)\delta b_{j}(0)\rangle_{\mathrm{QLT}}\simeq \int\dif\mathbf{k}~P_{ij}(\mathbf{k})\frac{\sin{(kct)}}{kct},
\end{equation}
which is shown as the dashed-dotted curve in Fig.~\ref{fig:phi}. It is observed to capture roughly the non-zero range of $\varphi(t)$. It fails, however, to reproduce the slope of the fall-off at small times, which is one critical feature. An improved approximation beyond the QLT (bQLT), proposed in~\cite{PhysRevD.65.023002}, consists in estimating $\langle e^{i\mathbf{k}\cdot\mathbf{x}(t)}\rangle$ in the white-noise limit:
\begin{eqnarray}
    \label{eqn:dbdb_bqlt}
    \langle \delta b_{i}(t)\delta b_{j}(0)\rangle_{\mathrm{bQLT}}&\simeq& \nonumber\\ 
    && \hspace{-2.5cm}\int\dif\mathbf{k}~P_{ij}(\mathbf{k})\exp{\left(-\frac{1}{3}k^2c^2t\int_0^t\dif t'C(t')\right)},
\end{eqnarray}
where $C(t')$ is the auto-correlation function of the pitch-angle cosine, which relies on the desired expression of $\langle v_{0i}v_{i}(t')\rangle$. In this framework, the difficulty is thus to face a highly non-linear equation for $u(t)$ that iterative methods might solve.

Rather than adopting such numerically-demanding iterative methods, the approach followed in this study consists in approximating $\varphi(t)$ with an exponential function, 
\begin{equation}
    \label{eqn:rednoise}
    \varphi(t)\simeq\exp{(-t/\tau)},
\end{equation}
which corresponds to a red-noise approximation. Just as in the case of the white-noise limit $\varphi_{\mathrm{WN}}(t)=\tau\delta(t)$, this approximation requires the introduction of a time-scale parameter $\tau$, related to the correlation time scale of the turbulence. By essence, the slope of the fall-off of $\varphi(t)$ at small times is captured accurately, as shown by the red dashed curve in Fig.~\ref{fig:phi}. Compared to the white-noise limit, the use of the red-noise approximation allows the parameter $\tau$ to come into play to shape the memory of the time scale over which the particles experience magnetic field values correlated to the initial ones. The introduction of such a memory in the process is essential to reproduce, for $\rho\lesssim \pi L_{\mathrm{c}}/L_{\mathrm{max}} (\simeq 0.2)$, the ascent that follows the rapid fall off in the time evolution of the velocity auto-correlations of the particles. This ascent, together with the final fall off, will be shown to be well reproduced in~\ref{subsec:kraichnan}, validating hence the red-noise approximation despite the overestimation of $\varphi(t)$ for $t\gtrsim\tau$. Finally, another benefit of this approach is, as explicited below, the possibility to carry out analytically in the Laplace reciprocal space a partial summation of the Dyson series that converges to a physical solution.  

The correlation time scale parameter is determined, for $\rho\gtrsim \pi L_{\mathrm{c}}/L_{\mathrm{max}}$, as $\tau\simeq L_\mathrm{c}/c$. This is because in this rigidity regime, particles can travel over a distance $L_\mathrm{c}$ undergoing small deflections only. On the other hand, for $\rho \lesssim \pi L_{\mathrm{c}}/L_{\mathrm{max}}$, the following heuristic estimate, similar to that found in~\cite{PhysRevD.65.023002}, is observed to reproduce simulations:
\begin{equation}
    \label{eqn:tau}
    \tau\simeq\frac{1}{c}\frac{\int_{k_\star}^{k_{\mathrm{max}}}\dif k k^{-1}\mathcal{E}(k)}{\int_{k_\star}^{k_{\mathrm{max}}}\dif k \mathcal{E}(k)}.
\end{equation}
In this regime of rigidity, $\tau$ inherits a $\rho$ dependency from that of the lower boundary $k_\star(\rho)=\rho_\star k_{\mathrm{min}}/\rho$ with $\rho_\star=2L_{\mathrm{c}}/(\pi L_{\mathrm{max}})$. The truncation in the wavenumber integration range selects modes for which particles do not experience spiral motions around the corresponding large-scale magnetic field lines over several Larmor times, modes that hence prevent decorrelations from occurring on relevant time scales.

\subsection{Partial summation}
\label{subsec:kraichnan}

The aim is to carry out a summation of the infinite Dyson series for $u(t)$ (equation~\ref{eqn:udyson}) that provides us with a physical solution. Although this series is absolutely convergent for all $t$, the convergence for $t>3/(2\delta\Omega^2\tau)$ requires very many terms to be accounted for, which is rapidly challenging numerically. In addition, any truncation after any finite number of terms implies $u(t)<0$ or $u(t)\rightarrow\infty$ for $t\rightarrow\infty$ (see Appendix~\ref{sec:app}), which is physically unacceptable. It is therefore preferable to resort to partial summation schemes.

The simplest partial summation scheme is to substitute the full propagator for that of \cite{Bourret1962}, consisting in the sum of unconnected diagrams only:
\begin{equation}\label{eqn:bourret}
\begin{tikzpicture}
  \begin{feynman}
    \vertex (a1);
    \vertex [right=1.2cm of a1] (a2){~$\simeq$~};
    \vertex [right=0.4cm of a2] (a3);
    \vertex [right=1.2cm of a3] (a4){~+~};
    \vertex [right=0.4cm of a4] (a5);
    \vertex [right=1.2cm of a5] (a6);
    \vertex [right=1.2cm of a6] (a7);
    \vertex [right=1.2cm of a7] (a8){.};
    \diagram*{
      (a1) --[double,double distance=0.3ex,thick] (a2),
      (a3) --[plain] (a4),
      (a5) --[plain] (a6) --[plain] (a7)  --[double,double distance=0.3ex,thick]  (a8),
      (a6) -- [scalar, out=90, in=90, looseness=1.5,thick] (a7)
    };
  \end{feynman}
\end{tikzpicture}
\end{equation}
This partial summation is the exact solution in the white-noise limit due to the cancellation of all crossed or nested diagrams. However, it proves to be insufficient in the case of red noise. A Laplace transform of this equation allows for transforming the integral equation for $u^{(1)}(t)$ into a linear one for $U^{(1)}(p)$, which, after an inverse transform, leads to
\begin{equation}
\label{eqn:u1}
  u^{(1)}(t)=\frac{A}{2B}e^{-At/6\tau}(e^{Ct/\tau}-1),
\end{equation}
with $A=3+\sqrt{9-24\tau^2\delta\Omega^2}$, $B=A-3$, and $C=\sqrt{1-8\tau^2 \delta\Omega^2/3}$. This solution gives rise to nonphysical oscillations around 0 for $t>3/(2\delta\Omega^2\tau)$. 

The next partial summation scheme, known as the propagator of \cite{Kraichnan:1961:DNS}, consists in substituting the free propagator inside the dotted loop for the full propagator so as to sum all unconnected and nested diagrams:
\begin{equation}\label{eqn:kraichnan}
\begin{tikzpicture}
  \begin{feynman}
    \vertex (a1);
    \vertex [right=1.2cm of a1] (a2){~$\simeq$~};
    \vertex [right=0.4cm of a2] (a3);
    \vertex [right=1.2cm of a3] (a4){~+~};
    \vertex [right=0.4cm of a4] (a5);
    \vertex [right=1.2cm of a5] (a6);
    \vertex [right=1.2cm of a6] (a7);
    \vertex [right=1.2cm of a7] (a8){.};
    \diagram*{
      (a1) --[double,double distance=0.3ex,thick] (a2),
      (a3) --[plain] (a4),
      (a5) --[plain] (a6) --[double,double distance=0.3ex,thick] (a7)  --[double,double distance=0.3ex,thick]  (a8),
      (a6) -- [scalar, out=90, in=90, looseness=1.5,thick] (a7),
    };
  \end{feynman}
\end{tikzpicture}
\end{equation}
\begin{figure}[t]
\centering
\includegraphics[width=\columnwidth]{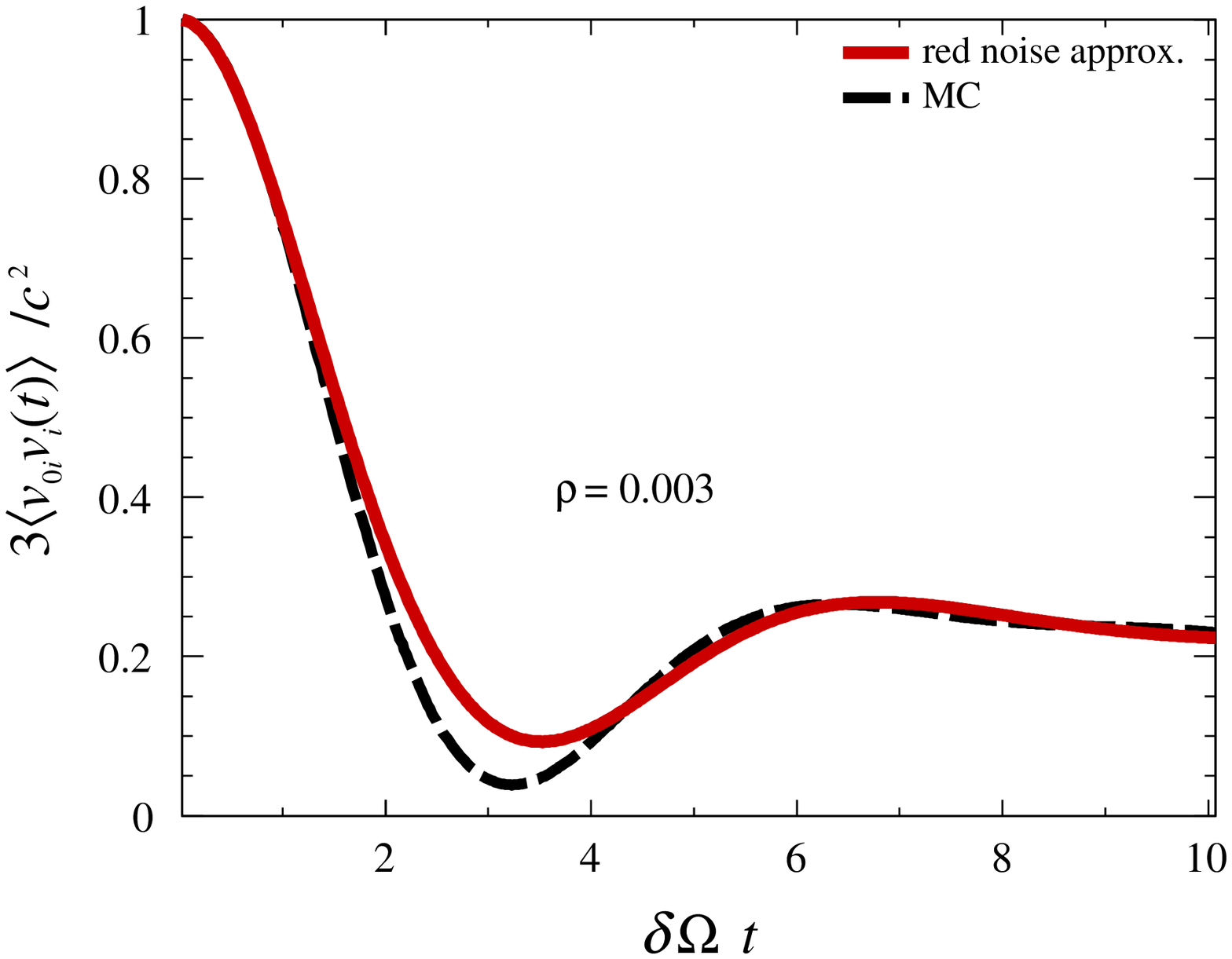}
\includegraphics[width=\columnwidth]{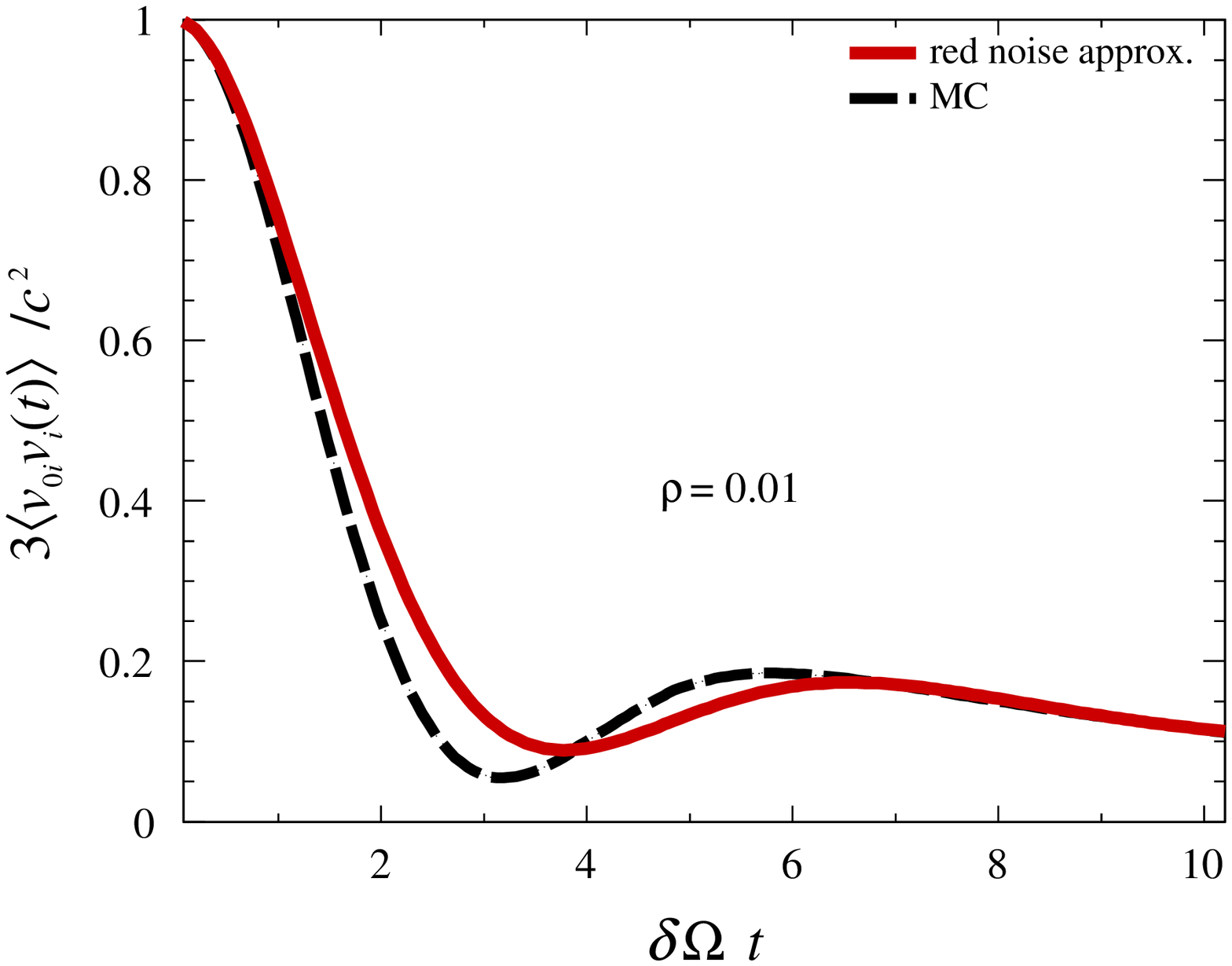}
\includegraphics[width=\columnwidth]{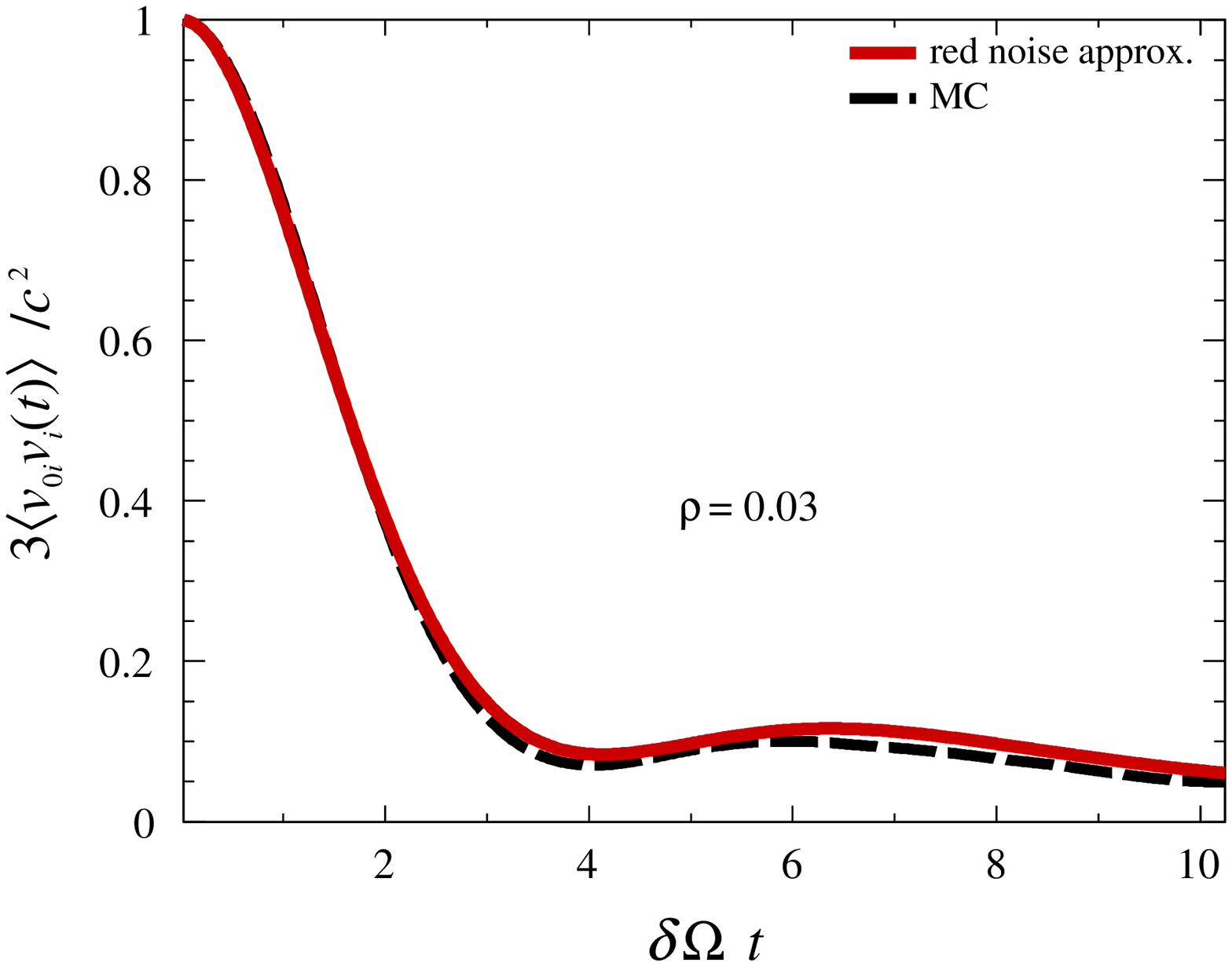}
\caption{Time dependence of the auto-correlation of the particle velocities expressed in units of $c^2/3$ ($u(t)$ function) for values of rigidities $\rho=0.003$ (top), $\rho=0.01$ (middle), and $\rho=0.03$ (bottom).}
\label{fig:R1}
\end{figure}
\begin{figure}[t]
\centering
\includegraphics[width=\columnwidth]{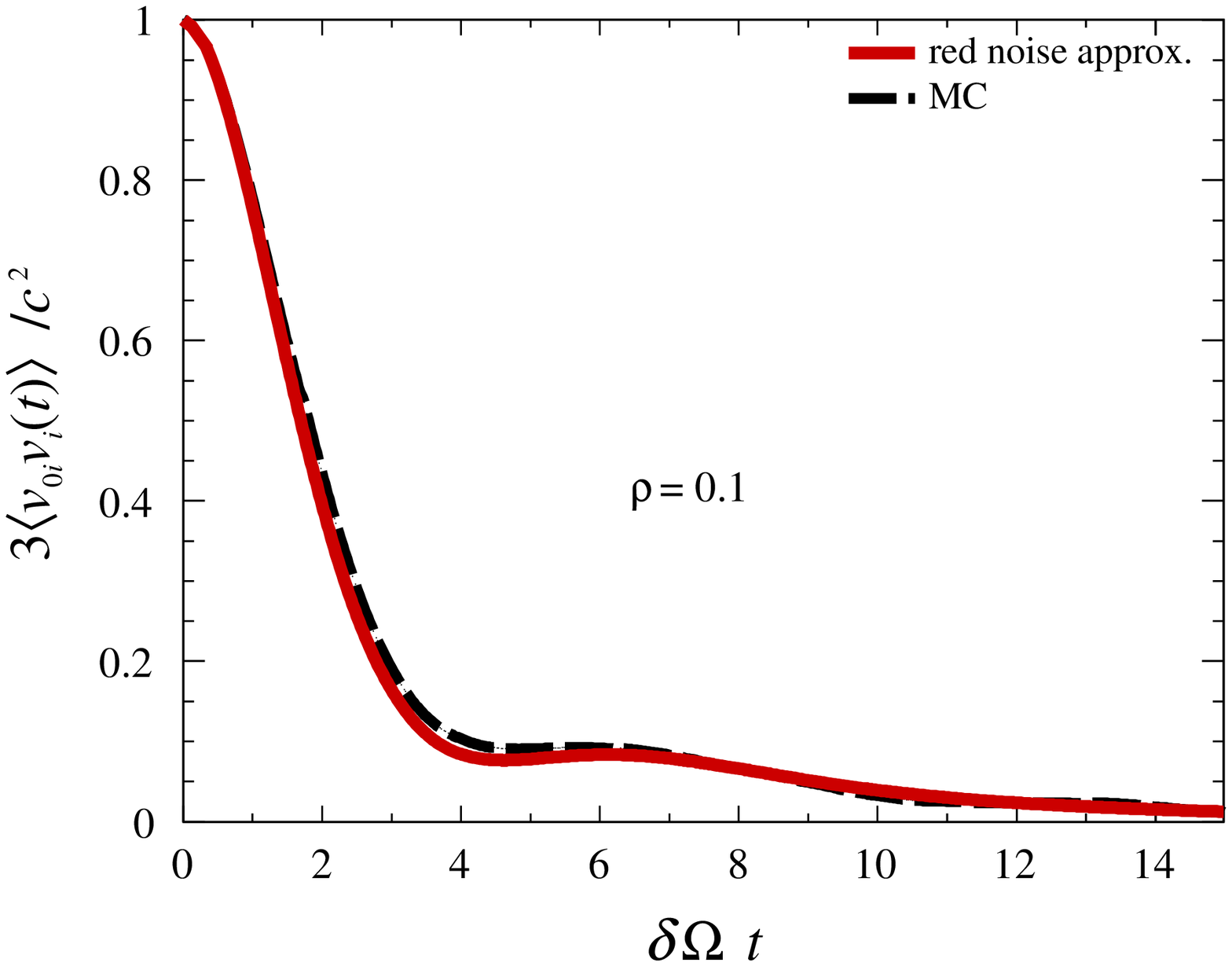}
\includegraphics[width=\columnwidth]{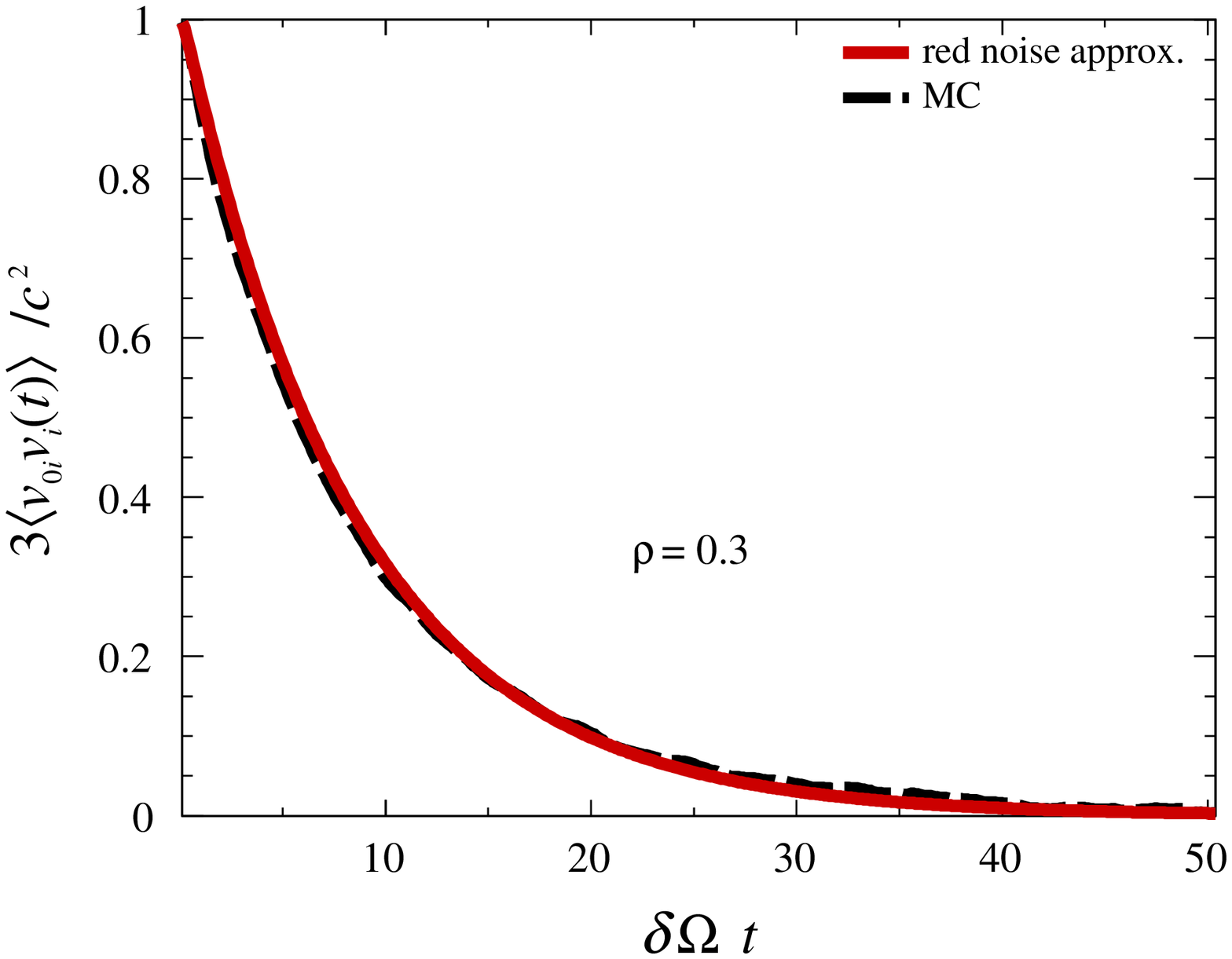}
\includegraphics[width=\columnwidth]{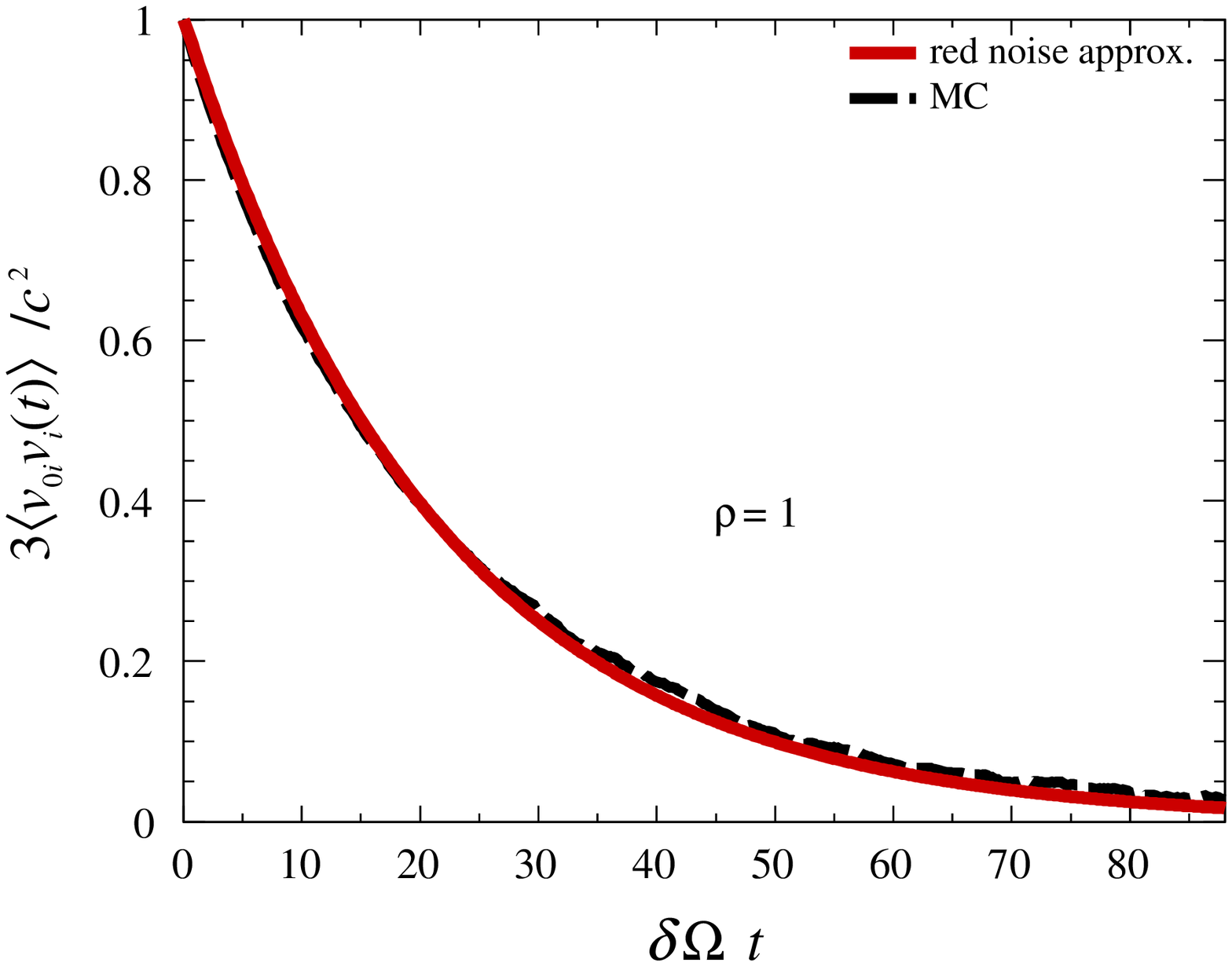}
\caption{Same as Fig.~\ref{fig:R1} for values of rigidities $\rho=0.1$ (top), $\rho=0.3$ (middle), and $\rho=1$ (bottom).}
\label{fig:R2}
\end{figure}
This is a non-linear equation in the time domain, which is decoupling in the Laplace reciprocal space so that
\begin{eqnarray}
\label{eqn:U2}
  \left[U^{(2)}(p)\right]^{-1}&=&p-\sum_{n\geq 1}\frac{\left(-2~\delta\Omega^2/3\right)^{n}}{(p+\tau^{-1})^n(p+2\tau^{-1})^{n-1}} \nonumber \\
  && \hspace{-2cm}-\sum_{n\geq 3}\frac{\left(-2~\delta\Omega^2/3\right)^{n}}{(p+\tau^{-1})^2(p+2\tau^{-1})^{n-1}(p+3\tau^{-1})^{n-2}}\nonumber \\
  && \hspace{-2cm}-\sum_{n\geq 4}\frac{\left(-2~\delta\Omega^2/3\right)^{n}}{(p+\tau^{-1})^2(p+2\tau^{-1})^{2}(p+3\tau^{-1})^{n-2}(p+4\tau^{-1})^{n-3}}\nonumber \\
  &&-~\cdots
\end{eqnarray}
The solution $u^{(2)}(t)$ is then obtained by making use of the numerical Stehfest scheme of the inverse Laplace transform. Although $n$ must be formally sent to infinity, a truncation to $n\leq 2$ of this partial summation turns out to provide satisfactory results. Such a truncation corresponds to approximating equation~\ref{eqn:kraichnan} by
\begin{equation}\label{eqn:kraichnansimplified}
\begin{tikzpicture}
  \begin{feynman}
    \vertex (a1);
    \vertex [right=1.2cm of a1] (a2){~$\simeq$~};
    \vertex [right=0.4cm of a2] (a3);
    \vertex [right=1.2cm of a3] (a4){~+~};
    \vertex [right=0.4cm of a4] (a5);
    \vertex [right=1.2cm of a5] (a6);
    \vertex [right=1.2cm of a6] (a7);
    \vertex [right=1.2cm of a7] (a8){~+~};
    \vertex [below=of a4] (a9);
    \vertex [right=1.cm of a9] (a10);
    \vertex [right=0.3cm of a10] (a11);
    \vertex [right=1.cm of a11] (a12);
    \vertex [right=0.3cm of a12] (a13);
    \vertex [right=1.cm of a13] (a14){~.~};
    \diagram*{
      (a1) --[double,double distance=0.3ex,thick] (a2),
      (a3) --[plain] (a4),
      (a5) --[plain] (a6) --[plain] (a7)  --[double,double distance=0.3ex,thick]  (a8),
      (a6) -- [scalar, out=90, in=90, looseness=1.5,thick] (a7),
      (a9) --[plain] (a10) --[plain] (a11) --[plain] (a12) --[plain] (a13) --[double,double distance=0.3ex,thick] (a14),
      (a10) -- [scalar, out=90, in=90, looseness=1.5,thick] (a13),
      (a11) -- [scalar, out=90, in=90, looseness=1.5,thick] (a12),

    };
  \end{feynman}
\end{tikzpicture}
\end{equation}

Comparisons of $u^{(2)}(t)$ with the Monte-Carlo results are shown in Fig.~\ref{fig:R1} and~~\ref{fig:R2} for several values of rigidities. For $\rho\leq 0.01$, a range of rigidities such that the Larmor radius of the particles is smaller than the smallest wavelengths of the turbulence, the main features of the time evolution of $u(t)$ are captured by the various approximations leading to $u^{(2)}(t)$ but a good agreement cannot be claimed. By contrast, for values of $\rho>0.01$ such that the Larmor radius is in gyro-resonance with wavelengths of the turbulence or larger than the size of the largest eddies, agreement between the red-noise approximation and the Monte-Carlo is observed. In particular, the non-exponential fall-off that holds in the gyro-resonant regime, probed in several numerical studies~\citep{Candia:2004yz,Fraschetti:2012cm}, is reproduced. As $\rho$ is increasing, the range over which $u^{(2)}(t)$ is significantly non-zero (scattering time scale) increases from several to very many multiples of $2\pi$ in terms of $\delta\Omega~t$. This illustrates that in the high energy regime, the scattering time scale gets much larger than the correlation one; an exponential fall-off is then recovered~\citep{Plotnikov:2011me}.


\section{Hints in presence of a mean field}
\label{sec:meanfield}

\begin{figure}[thp]
\centering
\includegraphics[width=\columnwidth]{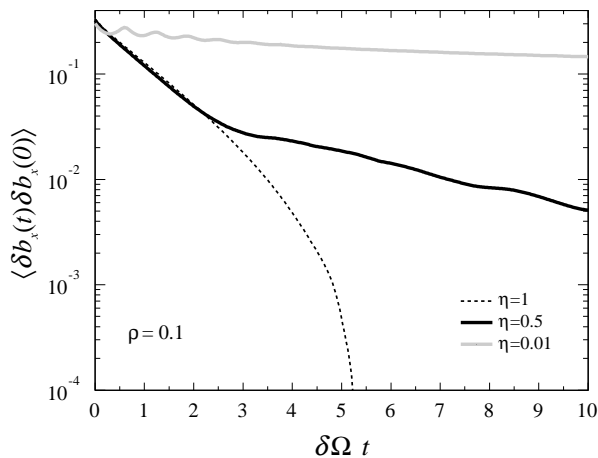}
\caption{Time dependence of the expectation value of $\langle \delta b_x(t)\delta b_x(0)\rangle$ for three different turbulence levels and $\rho=0.1$.}
\label{fig:dbxdbx}
\end{figure}

Although the case of a purely turbulent magnetic field is relevant in some astrophysical contexts, most cases of interest must deal with intermediate or low levels of turbulence. The numerous Monte-Carlo studies exploring the $(\eta,\rho)$ parameter space have revealed non-trivial behaviors, in particular for the running perpendicular and anti-symmetric coefficients. Adapting the formalism presented in Section~\ref{sec:acmodel} to the case of a mean field on top of a turbulence turns out to be non-trivial without introducing several free parameters. In the following, only a few hints illustrating the required additional ingredients are given. 

One key ingredient is the modeling of the 2-pt correlation function of the magnetic field experienced by the particles between two successive times. The presence of a mean field modifies the picture presented in Section~\ref{subsec:dbdb} in a way that the red-noise approximation is no longer valid. This is illustrated in Fig.~\ref{fig:dbxdbx}, where the $xx$-component of the $\langle \delta b_i(t)\delta b_j(0)\rangle$ function is shown for $\rho=0.1$ (gyro-resonant regime) and for three levels of turbulence, namely $\eta=1$, $\eta=0.5$, and $\eta=0.01$. In the cases $\eta<1$, the mean field is oriented in the $z$ direction, perpendicular to the initial directions of the test particles. The mean field is observed to introduce a longer time-scale memory on top of that stemming from the turbulence, the intensity of which relative to that of the turbulence is increasing with decreasing values of $\eta$.  

The longer time scale memory is expected to induce oscillations in the velocity auto-correlation function on a shorter time scale compared to that of the mean field-induced oscillations, which are driven by the Larmor frequency of the particles in spiral motion around the mean field. This effect has been observed in Monte-Carlo simulations to give rise to a sub-diffusive regime for the perpendicular and anti-symmetric running coefficients prior to reaching the plateau of the diffusion regime~\citep[e.g.][]{Candia:2004yz,PhysRevD.65.023002,Fraschetti:2012cm}. A comprehensive characterization of the $\langle \delta b_i(t)\delta b_j(0)\rangle$ function is however beyond the scope of this study and is left for a future one. 

\section{Discussion} 
\label{sec:discussion}

A derivation of the diffusion coefficient describing the propagation of cosmic rays in a 3D isotropic turbulence has been presented, extending the pioneering work of \cite{Plotnikov:2011me} to the case of a range of rigidities gyro-resonant with the power spectrum of the turbulence. The derivation relies on a single time parameter related to the turbulence correlation time that describes in an economic way the 2-pt correlation function of the magnetic field experienced by the particles between two successive times as a red-noise process. Technically, such a red-noise approximation makes possible a partial re-summation of the Dyson series that solves the equation of motion for the particle velocities. 
 
The red-noise approximation has shown to be a valid one for both the gyro-resonant and the high-rigidity regimes. However, it fails to describe the regime in which particles have a Larmor radius smaller than the smallest scale of the turbulence. In such cases, a modeling of the $\langle \delta b_i(t)\delta b_j(0)\rangle$ functions beyond an exponential fall-off together with a summation of the Dyson series beyond the approximations used in Section~\ref{subsec:kraichnan} is required.

More generally, better modelings of the $\langle \delta b_i(t)\delta b_j(0)\rangle$ functions are necessary in the case of the presence of a mean field. The kind of formalism presented in this study could then be used to infer the various dependencies of the three coefficients in equation~\ref{eqn:Dij} in the whole parameter space $(\rho,\eta)$. 
 
\acknowledgments
I thank Haris Lyberis for his numerous works at an earlier stage of this study, and Carola Dobrigkeit for her careful reading of the paper. 

\appendix

\section{Non-convergent truncation of the Dyson series}
\label{sec:app}

In the same spirit as the examples presented in~\cite{Kraichnan:1961:DNS}, the calculation carried out in this appendix is an illustration of the benefit to restrict the summation of the Dyson series to tractable classes of terms to all orders such as equation~\ref{eqn:kraichnan}. 

The diagrams, denoted as $f_{n}^{k}(t)$, are classified below accordingly to two numbers, $n=2m$ and $k=2\ell$, with $n$ the total number of points and $k$ the number of crossed or nested points. Diagrams with a different topology but sharing the same $k$ and $n$ numbers have equal contributions. For instance, the following nested and crossed diagrams are equal:
\begin{equation}\label{eqn:k2}
\begin{tikzpicture}
  \begin{feynman}
    \vertex (a1){~$f_{4}^{2}(t)=$~};
    \vertex [right=1.2cm of a1] (a2);
    \vertex [right=0.5cm of a2] (a3);
    \vertex [right=0.5cm of a3] (a4);
    \vertex [right=0.5cm of a4] (a5);
    \vertex [right=0.5cm of a5] (a6){~=~};
    \vertex [right=0.5cm of a6] (a7);
    \vertex [right=0.5cm of a7] (a8);
    \vertex [right=0.5cm of a8] (a9);
    \vertex [right=0.5cm of a9] (a10);
    \vertex [right=0.5cm of a10] (a11);
    \vertex [right=0.5cm of a11] (a12){,};
    \diagram*{
      (a1) --[plain] (a6),
      (a2) -- [scalar, out=90, in=90, looseness=1.5,thick] (a5),
      (a3) -- [scalar, out=90, in=90, looseness=2,thick] (a4),
      (a7) --[plain] (a12),
      (a8) -- [scalar, out=90, in=90, looseness=2,thick] (a10),
      (a9) -- [scalar, out=90, in=90, looseness=2,thick] (a11),
    };
  \end{feynman}
\end{tikzpicture}
\end{equation}

\begin{equation}\label{eqn:k4}
\begin{tikzpicture}
  \begin{feynman}
    \vertex (a1){~$f_{6}^{4}(t)=$~};
    \vertex [right=1.2cm of a1] (a2);
    \vertex [right=0.5cm of a2] (a3);
    \vertex [right=0.5cm of a3] (a4);
    \vertex [right=0.5cm of a4] (a5);
    \vertex [right=0.5cm of a5] (a6);
    \vertex [right=0.5cm of a6] (a7);
    \vertex [right=0.5cm of a7] (a8){~=~};
    \vertex [right=0.5cm of a8] (a9);
    \vertex [right=0.5cm of a9] (a10);
    \vertex [right=0.5cm of a10] (a11);
    \vertex [right=0.5cm of a11] (a12);
    \vertex [right=0.5cm of a12] (a13);
    \vertex [right=0.5cm of a13] (a14);
    \vertex [right=0.5cm of a14] (a15);
    \vertex [right=0.5cm of a15] (a16){~=~};
    \vertex [right=0.5cm of a16] (a17);
    \vertex [right=0.5cm of a17] (a18);
    \vertex [right=0.5cm of a18] (a19);
    \vertex [right=0.5cm of a19] (a20);
    \vertex [right=0.5cm of a20] (a21);
    \vertex [right=0.5cm of a21] (a22);
    \vertex [right=0.5cm of a22] (a23);
    \vertex [right=0.5cm of a23] (a24){~=~};
    \vertex [right=0.5cm of a24] (a25);
    \vertex [right=0.5cm of a25] (a26);
    \vertex [right=0.5cm of a26] (a27);
    \vertex [right=0.5cm of a27] (a28);
    \vertex [right=0.5cm of a28] (a29);
    \vertex [right=0.5cm of a29] (a30);
    \vertex [right=0.5cm of a30] (a31);
    \vertex [right=0.5cm of a31] (a32){~.~};
    \diagram*{
      (a1) --[plain] (a8),
      (a2) -- [scalar, out=90, in=90, looseness=1.,thick] (a7),
      (a3) -- [scalar, out=90, in=90, looseness=2,thick] (a4),
      (a5) -- [scalar, out=90, in=90, looseness=2,thick] (a6),
      (a9) --[plain] (a16),
      (a10) -- [scalar, out=90, in=90, looseness=1.2,thick] (a14),
      (a11) -- [scalar, out=90, in=90, looseness=2,thick] (a12),
      (a13) -- [scalar, out=90, in=90, looseness=1.4,thick] (a15),
      (a17) --[plain] (a24),
      (a19) -- [scalar, out=90, in=90, looseness=1.2,thick] (a23),
      (a21) -- [scalar, out=90, in=90, looseness=2,thick] (a22),
      (a18) -- [scalar, out=90, in=90, looseness=1.4,thick] (a20),
      (a25) --[plain] (a32),
      (a26) -- [scalar, out=90, in=90, looseness=1.5,thick] (a28),
      (a27) -- [scalar, out=90, in=90, looseness=1.5,thick] (a30),
      (a29) -- [scalar, out=90, in=90, looseness=1.5,thick] (a31),
    };
  \end{feynman}
\end{tikzpicture}
\end{equation}
We remind that dashed lines indicate integrations over the ordered times crossing the continuous line. For instance, 
\begin{equation}\label{eqn:ex}
\begin{tikzpicture}
  \begin{feynman}
    \vertex (a24);
    \vertex [right=0.5cm of a24] (a25);
    \vertex [right=0.5cm of a25] (a26);
    \vertex [right=0.5cm of a26] (a27);
    \vertex [right=0.5cm of a27] (a28);
    \vertex [right=0.5cm of a28] (a29);
    \vertex [right=0.5cm of a29] (a30);
    \vertex [right=0.5cm of a30] (a31);
    \vertex [right=0.5cm of a31] (a32){~=~};
    \diagram*{
      (a25) --[plain] (a32),
      (a26) -- [scalar, out=90, in=90, looseness=1.5,thick] (a28),
      (a27) -- [scalar, out=90, in=90, looseness=1.5,thick] (a30),
      (a29) -- [scalar, out=90, in=90, looseness=1.5,thick] (a31),
    };
  \end{feynman}
\end{tikzpicture}
\left(\frac{-2\delta\Omega^2}{3}\right)^{3}\int_0^t\dif t_1\int_0^{t_1}\dif t_2\int_0^{t_2}\dif t_3\int_0^{t_3}\dif t_4\int_0^{t_4}\dif t_5\int_0^{t_5}\dif t_6~\varphi(t_1-t_3)\varphi(t_2-t_5)\varphi(t_4-t_6).
\end{equation}

An attempt to carry out a summation of all terms up to some order $k_{\mathrm{max}}$ can be made by weighting all $f_{2m}^{2\ell}$ functions by the combinatorics that determines their number of occurrence. For $k=0$, one is left with the series of unconnected diagrams, each of them with weight 1; this is the Bourret propagator. For $k=2$, there are $2(m-1)$ ways to insert one nested or crossed diagram among $2m-4$ unconnected ones. For $k=4$, a minimal number of $n=6$ points is required. There are four crossed/nested ways to join the points that give rise to the same contribution (equation~\ref{eqn:k4}), and, for $m>2$, there are $\binom{m-2}{m-3}$ distinct ways to insert any of these diagrams among $n-6$ points joined with unconnected diagrams. In addition, starting from $n=8$ points, there are four possibilities to connect any of the diagrams of equation~\ref{eqn:k2} into $n=8$ points, and there are $\binom{m-2}{2}$ ways to insert them among $n-8$ points. This leads to the contribution
\begin{equation}
u(t;k=4)=\sum_m \left[4\binom{m-2}{m-3}f_{6,m>2}^4(t)+4\binom{m-2}{2}f_{8,m>3}^4(t)\right].
\end{equation}
For $k=6$, a minimal number of $n=6$ points is required. There are six crossed/nested ways to join the points that give rise to the same contribution, $f_6^6(t)$. Contributions to $f_n^6(t)$ can also arise with $n=8$ (eight different diagrams), with $n=10$ by connecting $f_6^4(t)$ and $f_4^2(t)$ with any number of unconnected diagrams (sixteen different diagrams denoted as $f^{4+2}_{10}(t)$), and with $n=12$ by connecting three $f_6^4(t)$ diagrams (eight different diagrams). The associated combinatorics reads as
\begin{equation}
u(t;k=6)=\sum_m \left[6\binom{m-2}{m-3}f_{6,m>2}^6(t)+8\binom{m-3}{m-4}f_{8,m>3}^6(t)+16\binom{m-3}{m-5}f_{10,m>4}^{4+2}(t)+8\binom{m-3}{3}f_{12,m>5}^{2+2+2}(t)\right].
\end{equation}
The reasoning can be repeated for higher values of $k$, although the increasing number of terms gets quickly non-tractable. For $k=8$, the contribution reads as
\begin{eqnarray}
u(t;k=8)=\sum_m \bigg[24\binom{m-3}{m-4}f_{8,m>3}^8(t)&+&16\binom{m-4}{m-5}f_{10,m>4}^8(t)+24\binom{m-3}{m-5}f_{10,m>4}^{6+2}(t)+16\binom{m-4}{m-6}f_{12,m>5}^{4+4}(t) \nonumber\\
&+&32\binom{m-4}{m-6}f_{12,m>5}^{6+2}(t)+48\binom{m-4}{m-7}f_{14,m>6}^{4+2+2}(t)+16\binom{m-4}{4}f_{16,m>7}^{2+2+2+2}(t)\bigg],
\end{eqnarray}
and so on and so forth.

\begin{figure}[ht]
\centering
\includegraphics[width=0.6\columnwidth]{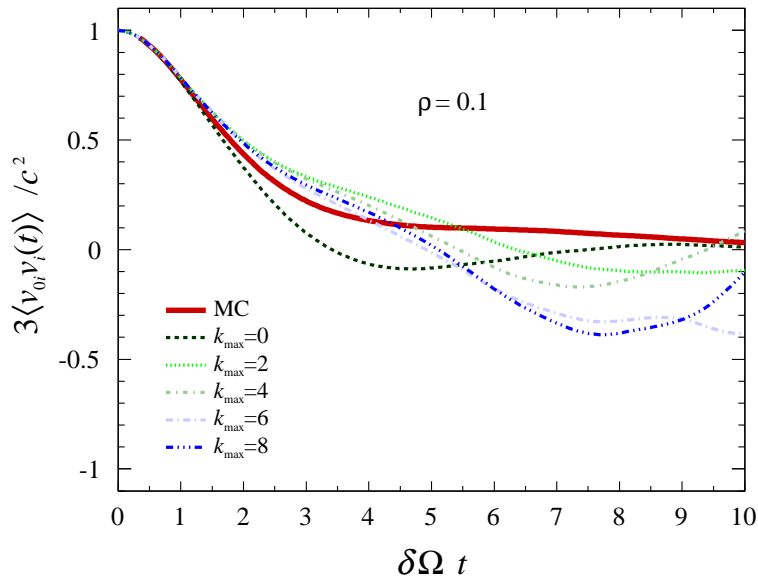}
\caption{$u(t;k_\mathrm{max})$ for several values of truncation $k_\mathrm{max}$.}
\label{fig:u}
\end{figure}

Comparisons between $u(t;k_\mathrm{max})=\sum_{k=0}^{k_\mathrm{max}}u(t;k)$ and $u^{(2)}(t)$ obtained from the Kraichnan propagator are shown in Fig.~\ref{fig:u}. It can be seen that the truncation to any value of $k_\mathrm{max}$ as high as $8$ fails to produce a physical solution for $t \gtrsim 2/(3\delta\Omega^2\tau)$. This clearly illustrates the benefit of using the Kraichnan propagator.

\bibliographystyle{aasjournal.bst}
\bibliography{bibliography}

\end{document}